\documentclass{article}
\usepackage[a4paper,
	left=2.5cm,
	right=2.5cm,
	top=2.5cm,
	bottom=2.5cm,
	includehead,
	includefoot,
	heightrounded,
	footskip=1.25cm]{geometry}
\usepackage{bm}
\usepackage{graphicx}
\usepackage{helvet}
\usepackage{caption}
\usepackage{subcaption}
\usepackage[backend=biber, bibstyle=numeric, doi=true, isbn=false, url=true, sorting=nyt, natbib=true, maxcitenames=5, maxbibnames=99, bibencoding=utf8, defernumbers=true, giveninits=true]{biblatex}
\usepackage{hyperref}
\usepackage[mode=buildnew]{standalone}
\usepackage{authblk}
\usepackage[frozencache=true,cachedir=minted-cache]{minted}
\usepackage{tikz}

\DeclareSourcemap{
	\maps[datatype=bibtex, overwrite]{
		\map{
			\pertype{article, inproceedings}
			\step[fieldset=publisher, null]
		}
		\map{
			\step[fieldset=address, null]
			\step[fieldset=location, null]
			\step[fieldset=note, null]
			\step[fieldset=series, null]
			\step[fieldset=url, null]
		}
	}
}
\makeatletter
\let\blx@rerun@biber\relax
\makeatother

\title{Off-By-One Implementation Error in J-UNIWARD}

\author{Benedikt Lorch\\Security and Privacy Lab\\Universität Innsbruck}

\date{May 31, 2023}

\begin{document}
	\maketitle

	\begin{abstract}
		J-UNIWARD is a popular steganography method for hiding secret messages in JPEG cover images. As a content-adaptive method, J-UNIWARD aims to embed into textured image regions where changes are difficult to detect. To this end, J-UNIWARD first assigns to each DCT coefficient an embedding cost calculated based on the image's Wavelet residual, and then uses a coding method that minimizes the cost while embedding the desired payload.

		Changing one DCT coefficient affects a $23 \times 23$ window of Wavelet coefficients. To speed up the costmap computation, the original implementation pre-computes the Wavelet residual and then considers per changed DCT coefficient a $23 \times 23$ window of the Wavelet residual. However, the implementation accesses a window accidentally shifted by one pixel to the bottom right.

		In this report, we evaluate the effect of this off-by-one error on the resulting costmaps. Some image blocks are over-priced while other image blocks are under-priced, but the difference is relatively small. The off-by-one error seems to make little difference for learning-based steganalysis.
	\end{abstract}

\section{Background}
	J-UNIWARD is a popular steganography method for hiding secret messages in JPEG cover images~\cite{Holub2014Uniward}. While embedding the desired payload, J-UNIWARD aims to minimize the distortion introduced by the embedding. The distortion is calculated as the relative sum of changes between cover image $\bm{X}$ and stego image $\bm{Y}$ after Wavelet filtering,
	
	\begin{equation}
		D(\bm{X}, \bm{Y}) = \sum\limits_{k=1}^3 \sum\limits_{u=1}^{n_1} \sum\limits_{v=1}^{n_2} \frac{\left| W_{uv}^{(k)}\left( \bm{X} \right) - W_{uv}^{(k)} \left( \bm{Y} \right) \right|}{\sigma + \left| W_{uv}^{(k)} \left(\bm{X} \right)\right|} \enspace.
		\label{eq:uniward_distortion}
	\end{equation}

\begin{figure}
		\centering
		\includegraphics{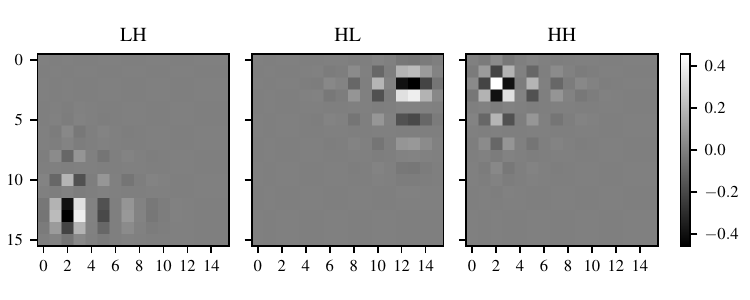}
		\caption{The three directional Wavelet filter kernels ($16 \times 16$) used by J-UNIWARD evaluate the smoothness along the horizontal~(LH), vertical~(HL), and diagonal~(HH) directions.}
		\label{fig:wavelet_filters}
	\end{figure}

	The Wavelet filter bank consists of three directional filters, which are shown in Fig.~\ref{fig:wavelet_filters}. The residuals are denoted as $W^{(k)}(\bm{X})$, where $k \in \{1, 2, 3\}$. The subscripts $u, v$ denote the spatial position and iterate from $1$ to the image height $n_1$ and image width $n_2$, respectively. The constant $\sigma$ controls the sensitivity of the distortion to the image content. The authors set it to $2^{-6}$~\cite{Holub2014Uniward}.
	
	The denominator in Eq.~\ref{eq:uniward_distortion} discourages embedding changes in regions where the cover's Wavelet residual is small, i.e., when the region is smooth in at least one direction.
	
	The distortion is defined in the spatial domain, but steganography with JPEG images embeds into their quantized DCT coefficients. To quantify the distortion introduced by changing quantized DCT coefficents, J-UNIWARD decompresses the quantized DCT coefficients and evaluates the distortion in the spatial domain.
	
	After computing the distortion for each embeddable element, the steganographer uses a coding method such as syndrome trellis codes~(STCs) to embed their secret message while minimizing the distortion.
	
\section{Original implementation}
	The steganographer computes for each DCT coefficient the distortion (aka cost) introduced by changing that coefficient. The result is also called costmap.
	A naive implementation would iterate over all DCT coefficients of the cover image $\bm{X}$, increment\footnote{Because of the absolute values in Eq.~\ref{eq:uniward_distortion}, the distortion of incrementing and decrementing the DCT coefficient is the same.} one DCT coefficient at a time, and evaluate the distortion between the cover and the modified cover $\bm{Y}$.

	The authors of J-UNIWARD published a Matlab and a C++ implementation.\footnote{\url{http://dde.binghamton.edu/download/stego_algorithms/}, accessed on 02.05.2023} Both implementations pre-compute several quantities to speed up the costmap computation.
	Note that numerator in Eq.~\ref{eq:uniward_distortion} is independent of the image content. Hence, the numerator can be pre-computed by first evaluating the spatial impact of changing one of the $8 \times 8$ DCT coefficients, and then evaluating the impact on the Wavelet residual. Changing one DCT coefficient impacts the whole $8 \times 8$ spatial block. Because the Wavelet filter kernels have size $16 \times 16$, changing one DCT coefficient impacts a block of $23 \times 23$ Wavelet residual coefficients.\footnote{$8 + 16 -1 = 23$} As a result, the implementation pre-computes for each of the three filter kernels a look-up table. Each of the three look-up tables contains how much changing one of the $64$ DCT coefficients impacts the $23 \times 23$ Wavelet residual coefficients.
	
	The denominator requires the Wavelet residual of the cover image, which needs to be computed only once.
	
	After pre-computing these quantities, the steganographer iterates over all DCT coefficients. The numerator can be read from the look-up table. For the denominator, the steganographer only needs to extract the right $23 \times 23$ window from the pre-computed Wavelet residual. Note that the $23 \times 23$ window in the denominator is the same for all DCT coefficients within the current $8 \times 8$ block.
	
\section{Off-by-one implementation error}

	\begin{figure}
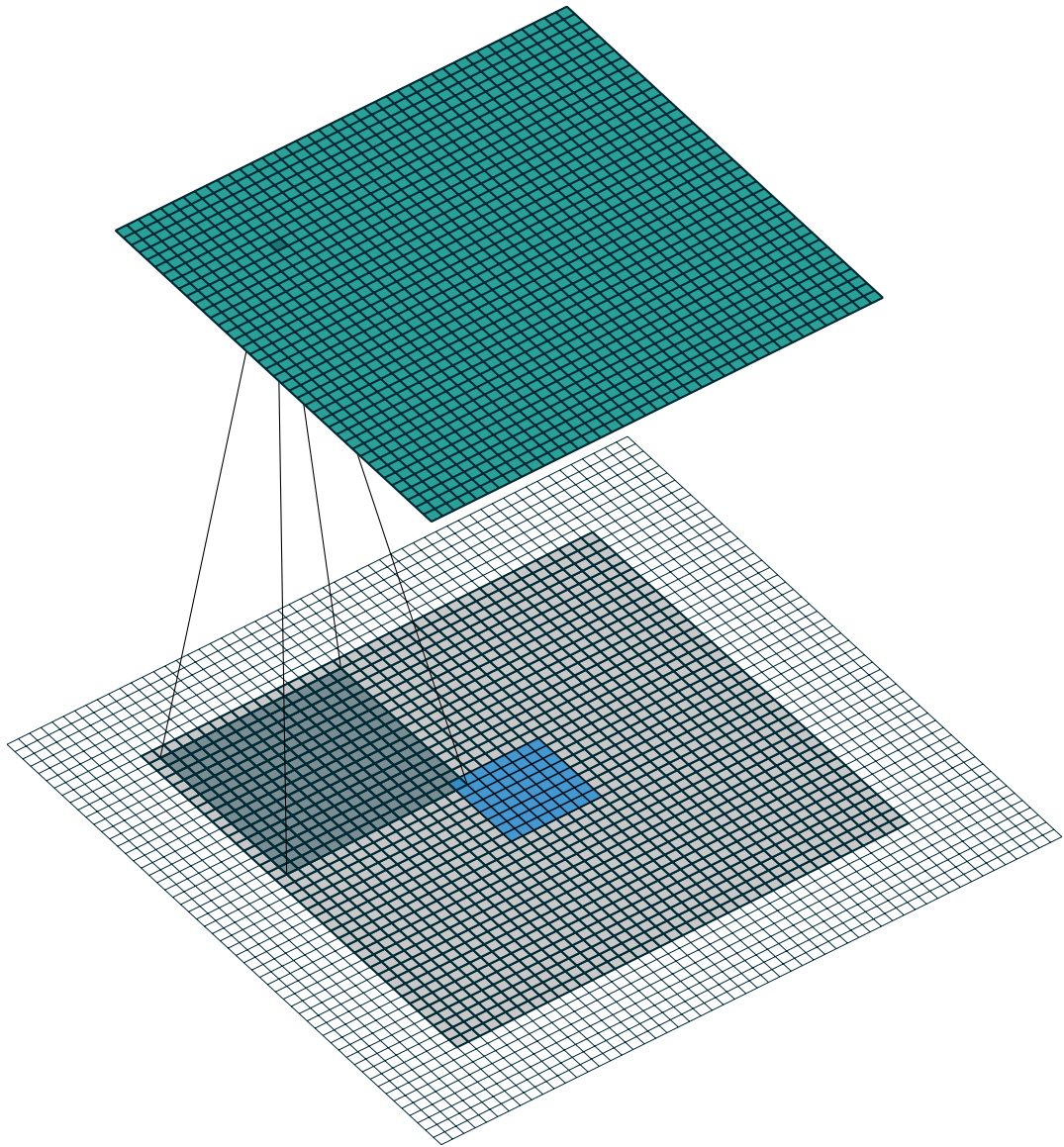

		\centering
		\includestandalone{images/filter_padding_illustration}
		\caption{The blue area shows an $8 \times 8$ cover image. The Wavelet residual (top green plane) is obtained as follows: (1)~The image is expanded by 16 pixels in all directions using symmetry padding (light gray area). (2)~The $16 \times 16$ filter kernel (dark gray) is slid over the image. Additional zero-padding is used to maintain the input resolution. (3)~The first position in the resulting Wavelet residual that is affected by the cover pixel $(0, 0)$ is located at the spatial offset $(8, 8)$, highlighted in dark green. However, the original implementation accesses the Wavelet residual starting from offset $(9, 9)$. The figure was created based on \url{https://github.com/vdumoulin/conv_arithmetic/}.}
		\label{fig:filter_padding_illustration}
	\end{figure}

	While extracting the $23 \times 23$ window from the pre-computed Wavelet residual, the implementation takes a window accidentally shifted by one pixel to the bottom right. The following steps describe the implementation error. For illustration purposes, we use a cover image of size $8 \times 8$, as shown in Fig.~\ref{fig:filter_padding_illustration}.

	\begin{enumerate}
		\item The cover image of size $8 \times 8$ is expanded by $16$ pixels in all directions using symmetry padding. The resulting image has size $40 \times 40$. The top left pixel is now at the spatial location $(16, 16)$ (0-based index).
		\item The $16 \times 16$ Wavelet filter kernel is slid over the image using \emph{same} mode and zero-padding. The resulting Wavelet residuals have the size $40 \times 40$. To obtain the same output resolution, Matlab's \texttt{imfilter} and scipy's \texttt{correlate2d} prepend $7$ zero pixels. Hence, the first output is computed from $7$ padded zeros to the top left and the top-left $9 \times 9$ values from the padded cover (indices 0 through 8). As a result, the first filter output that interacts with a ``real'' pixel value is at the output location $(8, 8)$.\footnote{While Matlab's \texttt{imfilter} expands the image by $7$ zeros to the top left to maintain the input resolution, note that the C++ implementation prepends $8$ zero row and columns.}
		\item When the cover pixel $(0, 0)$ is changed, this affects the Wavelet residual in the window $(8, 8)$ through $(30, 30)$ (upper bound is inclusive). However, the original implementation accesses the window $(9, 9)$ through $(31, 31)$.\footnote{This notation uses 0-based indexing. Add 1 to obtain 1-based indices as used by Matlab.}
	\end{enumerate}
	
	This off-by-one error is present both in the Matlab and the C++ implementation. The S-UNIWARD implementation is not affected because it does not do pre-computation. The implementations of side-informed UNIWARD~(SI-UNIWARD) and perturbed quantization~(PQ-UNIWARD) appear to contain the same error.

	While off-by-one errors often raise out-of-bounds access exceptions, such an exception does not occur here because the cover image was previously padded to avoid boundary artifacts.
	
	The Matlab implementation can be corrected as follows:
	\begin{minted}{matlab}
% Original Matlab implementation
subRows = row-modRow-6+padSize:row-modRow+16+padSize;
subCols = col-modCol-6+padSize:col-modCol+16+padSize;

% Suggested fix
subRows = row-modRow-7+padSize:row-modRow+15+padSize;
subCols = col-modCol-7+padSize:col-modCol+15+padSize;
	\end{minted}
	
	Similarly, the C++ implementation can be corrected as follows:
	\begin{minted}{cpp}
// Original C++ implementation
int subRowsFrom = row - modRow - 6 + config->padsize;
int subColsFrom = col - modCol - 6 + config->padsize;

// Suggested fix
int subRowsFrom = row - modRow - 7 + config->padsize;
int subColsFrom = col - modCol - 7 + config->padsize;
	\end{minted}
	
	A Python implementation of both the original and the corrected implementation is available here: \url{https://github.com/uibk-uncover/conseal}.

\section{Impact}
	We evaluate what impact this off-by-one error has, first on the costs per block, then on the embedding probabilities per DCT coefficient, and last for a steganalysis CNN.
	
\begin{figure}
		\centering
		\includegraphics{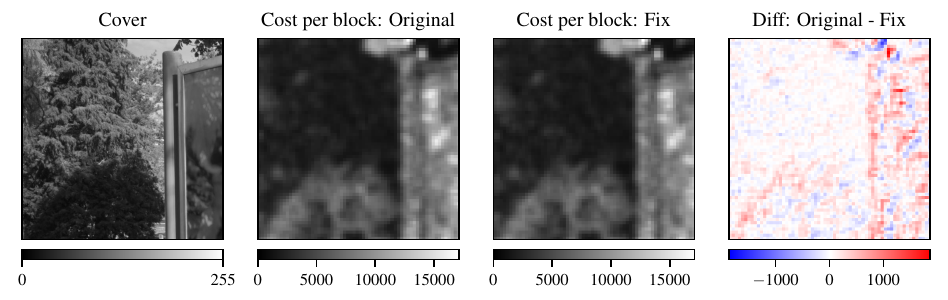}
		\caption{The left panel shows the cover image at JPEG quality 75. The second and third panels show the cost per block calculated with the original and the fixed implementation. The right panel shows the differences between the original and fixed cost per block. Negative values~(blue) mean that the original implementation under-prices the block; positive values~(red) show over-priced blocks.}
		\label{fig:original_fix_diff}
	\end{figure}

	\paragraph{Difference in block costs}
	Figure~\ref{fig:original_fix_diff} shows the J-UNIWARD cost per $8 \times 8$ block. The cost per block is calculated similar to Eq.~\ref{eq:uniward_distortion}, when the numerator is set to a constant $1$. The left panel in Fig.~\ref{fig:original_fix_diff} shows the cover image \emph{00001.jpg} from the ALASKA2 dataset at JPEG quality 75. The second panel shows the cost per block calculated with the original implementation. The third panel shows the cost per block calculated after fixing the off-by-one implementation error. The right panel shows the difference between the original costmap and the fixed costmap. Negative values~(blue) mean that the original implementation under-prices the block; positive values~(red) show over-priced blocks. It can be seen that the cost per block ranges between $0$ and $17000$, while the two costmaps differ in cost by less than $1500$.
	
\begin{figure}
		\centering
		\begin{subfigure}{.49\textwidth}
			\includegraphics{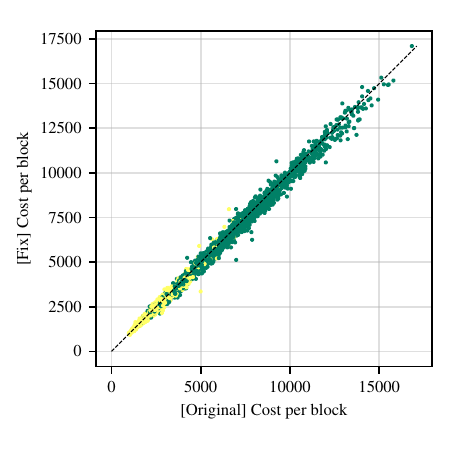}
			\caption{Difference in block costs between the original and the fixed implementation}
			\label{fig:scatter_cost_per_block}
		\end{subfigure}
		\begin{subfigure}{.49\textwidth}
			\includegraphics{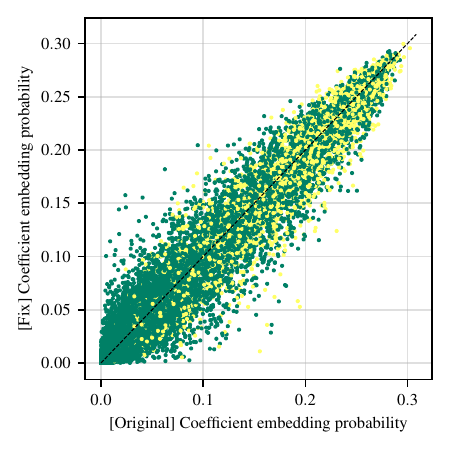}
			\caption{Difference in embedding probabilities between the original and the fixed implementation}
			\label{fig:scatter_embedding_probabilities}
		\end{subfigure}
		\caption{The two figures compare the original J-UNIWARD implementation and our variant that fixes the off-by-one error. The colors illustrate which blocks~(left) or DCT coefficients~(right) the embedding simulator randomly chose for embedding. As expected, the simulator prefers the blocks with low cost (left) and the coefficients with high probability (right). The simulator selects based on both cost and random numbers, which is why some blocks with higher cost and DCT coefficients with lower probability are sometimes chosen for embedding.}
		\label{fig:scatter_costs}
	\end{figure}

	Figure~\ref{fig:scatter_cost_per_block} again compares the two costmaps but only in one dimension. Each sample represents one $8 \times 8$ block. The x-coordinate shows the block cost calculated by the original implementation, the y-coordinate shows the block cost calculated by the fixed implementation. Of particular interest are samples above the diagonal, where the original implementation under-prices the block. However, the block costs are very similar between the two implementations, i.e., the points are all close to the diagonal.
	
	\paragraph{Difference in embedding probabilities}
	During the embedding, the costs are converted to embedding probabilities represented by a Gibbs distribution. The embedding probabilities for cover image \emph{00001.jpg} is shown in Figure~\ref{fig:scatter_embedding_probabilities}. Each sample represents one DCT coefficient. The x-coordinate shows the embedding probability calculated by the original implementation, the y-coordinate shows the embedding probability calculated by the fixed implementation. Compared to the block costs, the embedding probabilities deviate more from the diagonal. The spread probably comes from the exponential scaling in the Gibbs distribution. Nevertheless, the differences between the original and the fixed implementations are still small.

\begin{figure}
		\centering
		\begin{subfigure}{\textwidth}
			\includegraphics{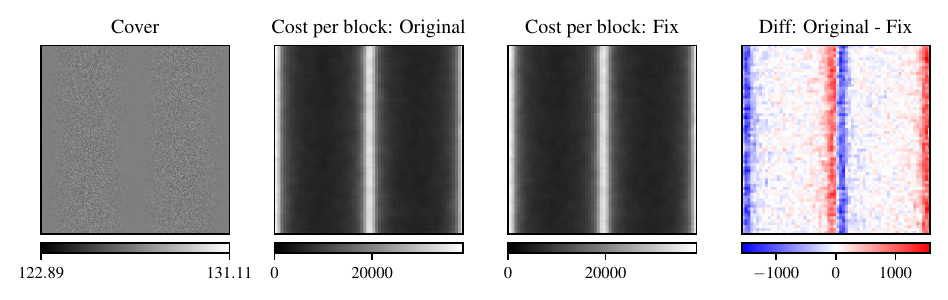}
			\caption{The synthetic cover alternates from left to right between smooth -- textured -- smooth -- textured -- smooth.}
			\label{fig:cost_per_block_synthetic_horizontal}
		\end{subfigure}
		\begin{subfigure}{\textwidth}
			\includegraphics{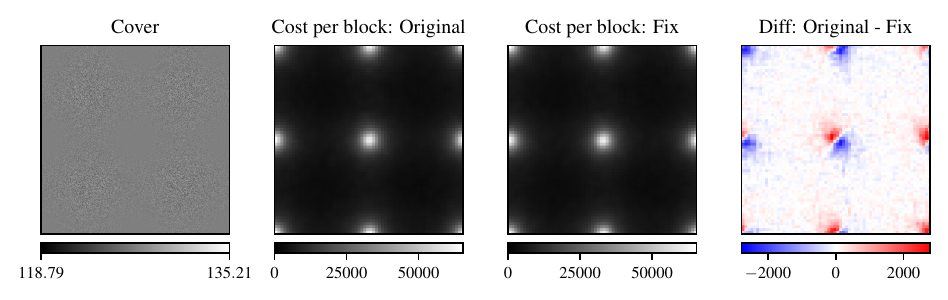}
			\caption{The synthetic cover alternates from left to right and top to bottom between smooth -- textured -- smooth -- textured -- smooth.}
			\label{fig:cost_per_block_synthetic_diagonal}
		\end{subfigure}
		\caption{The two synthetic images alternate between smooth and textured regions. Smooth regions have high cost, while textured regions have low cost. In transitions from smooth to textured regions, the original implementation under-prices the block cost (blue). In transitions from textured to smooth regions, the original implementation over-prices the block cost (red).}
		\label{fig:cost_per_block_synthetic}
	\end{figure}

	\paragraph{Influence of image content}
		We evaluate which image regions are particularly affected by the off-by-one error. For this purpose, we create synthetic images that alternate between smooth and textured regions. Two examples are shown in Fig.~\ref{fig:cost_per_block_synthetic}. The cover in the top row alternates from left to right. The cover in the bottom alternates from both left to right and top to bottom. As expected, smooth regions have high cost per block, while textured regions have low cost per block. In transitions from smooth to textured regions, the original implementation under-prices the block cost (blue). In transitions from textured to smooth regions, the original implementation over-prices the block cost (red). There is almost no difference in regions with a constant level of (non-)smoothness.

	\paragraph{Influence of JPEG quality}
		Overall, the costs per block increase for lower JPEG quality factors. We assume this is because removing high-frequency content makes the image appear more smooth, thereby increasing the cost of embedding. Lower quality factors seem to increase the relative cost difference between the two implementations, but the difference is still small. 

	\paragraph{Steganalysis}
		We trained an EfficientNet-B0 to distinguish cover images and stego images with JPEG quality 75. The luminance channel of the stego images was embedded with the original J-UNIWARD implementation and embedding rate $0.4$\,bpnzAC. The accuracy on in-distribution test images is $0.887$.
		
		We generated the same test images but with the fixed J-UNIWARD costmap implementation. The test accuracy is $0.885$. The barely noticeable difference in accuracy suggests that the off-by-one error has little impact.

		Experiments with JPEG quality 95 and 30 showed similar behavior.

\section*{References}
	\printbibliography[heading=none]

@article{Holub2014Uniward,
	doi = {10.1186/1687-417x-2014-1},
	url = {https://doi.org/10.1186/1687-417x-2014-1},
	year = {2014},
	month = jan,
	publisher = {Springer Science and Business Media {LLC}},
	volume = {2014},
	number = {1},
	author = {Vojt{\v{e}}ch Holub and Jessica Fridrich and Tom{\'{a}}{\v{s}} Denemark},
	title = {Universal distortion function for steganography in an arbitrary domain},
	journal = {{EURASIP} Journal on Information Security}
}

\end{document}